\documentclass{pasj00}
\draft

\begin{document}
\SetRunningHead{J. H. Park et al.}{The Light and Period Variations of the Eclipsing Binary BX Draconis}

\title{The Light and Period Variations of the Eclipsing Binary BX Draconis}

\author{Jang-Ho \textsc{Park}, Jae Woo \textsc{Lee}, Seung-Lee \textsc{Kim}, Chung-Uk \textsc{Lee}, and Young-Beom \textsc{Jeon}}
\affil{Korea Astronomy and Space Science Institute, Daejeon 305-348, Korea}
\email{pooh107162@kasi.re.kr, jwlee@kasi.re.kr, slkim@kasi.re.kr, leecu@kasi.re.kr, ybjeon@kasi.re.kr}


\maketitle

\begin{abstract}
New CCD photometric observations of BX Dra were obtained for 26 nights from 2009 April to 2010 June. The long-term photometric behaviors
of the system are presented from detailed studies of the period and light variations, based on the historical data and our new observations.
All available light curves display total eclipses at secondary minima and inverse O'Connell effects with Max I fainter than Max II,
which are satisfactorily modeled by adding the slightly time-varying hot spot on the primary star. A total of 87 times of minimum light
spanning over about 74 yrs, including our 22 timing measurements, were used for ephemeris computations. Detailed analysis of
the $O$--$C$ diagram showed that the orbital period has changed in combinations with an upward parabola and a sinusoidal variation.
The continuous period increase with a rate of $+$5.65$\times$10$^{-7}$ d yr$^{-1}$ is consistent with that calculated from
the Wilson-Devinney synthesis code. It can be interpreted as a mass transfer from the secondary to the primary star at a rate of
2.74$\times$10$^{-7}$ M$_\odot$ yr$^{-1}$, which is one of the largest rates for contact systems. The most likely explanation of
the sinusoidal variation with a period of 30.2 yrs and a semi-amplitude of 0.0062 d is a light-travel-time effect due to the existence
of a circumbinary object. We suggest that BX Dra is probably a triple system, consisting of a primary star with a spectral type of F0,
its secondary component of spectral type F1-2, and an unseen circumbinary object with a minimum mass of $M_3$ = 0.23 M$_\odot$.
\end{abstract}

\section{Introduction}

BX Dra (GSC 4192-0448, 2MASS J16061736+6245460, HIP 78891) was discovered to be a short-period variable by Strohmeier (1958) and classified
as an RR Lyr-type star with a period of 0.561192 days by Strohmeier et al. (1965). Some doubt about this classification and hints about
a binary nature were presented by Smith (1990), who proposed this star to be an ellipsoidal-type variable. Agerer \& Dahm (1995) reclassified
the variable as a $\beta$ Lyr-type eclipsing binary from their photographic and CCD photometry without any filter. They also determined
 new linear ephemeris from the CCD measurements and suggested that all available timings could be represented by a quadratic ephemeris,
in which the parabolic term (+5.56$\times$10$^{-10}$ d) indicates a continuous period increase. Pych et al. (2004) obtained
double-line radial-velocity curves of BX Dra with semi-amplitudes of $K_1$ = 80.0 km s$^{-1}$ and $K_2$ = 276.4 km s$^{-1}$ and found
that this system is an A-subtype contact binary with a spectral type of F0IV-V.

Recently, S\'anchez-Bajo et al. (2007, hearafter SGG), Kim et al. (2009, hearafter KIM), and Zola et al. (2010, hearafter ZOLA) made
CCD light curves in the $BVI$, $BV$, and $BVRI$ bandpasses, respectively. Assuming both components to have a convective envelope and
considering a cool spot on the primary component, SGG analyzed their light curves and found a third light contributing 2.4\% light
in the $B$ bandpass, 2.1\% in $V$, and 0.9\% in $I$. In addition, they computed the absolute dimensions of BX Dra
from their photometric parameters and the spectroscopic orbit of Pych et al. (2004) and suggested the third light may come from
a field object which is not bound to the eclipsing pair. On the other hand, ZOLA presented results of the modelling of
their multicolor light curves without considering either a third light source or a spot.

Although eclipsing minimum epochs have been reported assiduously by numerous workers, the period variation of BX Dra has not
yet been studied in detail. In this article, we present and discuss the long-term photometric behaviors of the binary system together
with the first detailed analysis of the $O$--$C$ diagram, based on our new CCD observations as well as the historical data.

\section{CCD Photometric Observations}

New CCD photometric observations of BX Dra were obtained during two observing seasons in 2009 April and between 2010 April and
June, using $BVR$ filters attached to the 61-cm reflector at Sobaeksan Optical Astronomy Observatory (SOAO) in Korea.
The observations of the 2009 season (SOAO09) were made on eight nights using a SITe 2K CCD camera, which has 2048$\times$2048 pixels and
an image field-of-view (FOV) of about 20$\arcmin$.5$\times$20$\arcmin.5$ at the f/13.5 Cassegrain focus of the telescope.
The observations of the 2010 season (SOAO10) were made on 18 nights using an FLI IMG4301E 2K CCD camera.
The new CCD chip has 2084$\times$2084 pixels and an FOV of about 20$\arcmin$.9$\times$20$\arcmin.9$.
The instruments and reduction methods were the same as those described by Lee et al. (2007, 2011). GSC 4192-0521 ($V$ = 11.0) and
GSC 4192-0617 ($V$ = 11.3) in the same observing field and with a color index similar to BX Dra were selected as comparison and
check stars, respectively. We detected no variations in the brightness difference between these two stars during our observing runs.
An observed image in SOAO09 is given as Figure 1, in which the comparison and check stars are marked C and K, respectively.
The 1$\sigma$ values of the dispersion of the (K$-$C) differences are about $\pm$0.009 mag in all bandpasses.

From the SOAO observations, we obtained 1620, 1622, and 1617 individual points in the $B$, $V$, and $R$ bandpasses, respectively, and
the sample is listed in Table 1. The light curves of BX Dra are plotted in Figure 2 as differential magnitude {\it versus} orbital phases,
which were computed according to the ephemeris for our hot-spot model described in section 3. The open circles and plus symbols are
the individual measures of the 2009 and 2010 seasons, respectively.

\section{Light-Curve Analysis And Spot Model}

As do historical light curves, our light curves of BX Dra show a flat bottom at the secondary minimum, which means that this system belongs
to the A-subtype of W UMa-type stars. In addition, the new data show the inverse O'Connell effect with Max I (at phase 0.25) fainter than
Max II (at phase 0.75) by about 0.014 and 0.008 mag for the $B$ and $V$, respectively, while the light levels in $R$ are equal at the quadratures.
These features usually indicate wavelength-dependent spot activity on the component stars.
In order to obtain a unique set of photometric solutions, we analyzed all available light curves (SGG, ZOLA, KIM, SOAO) using contact mode 3 of
the 2003 version of the Wilson-Devinney synthesis code (Wilson \& Devinney 1971, hereafter W-D). For this purpose, we normalized the level
at phase 0.75 and used a weighting scheme identical to that for the eclipsing binary GW Gem (Lee et al. 2009a). Table 2 lists the light-curve sets
for BX Dra analyzed in this article and the standard deviations ($\sigma$) of a single observation.

The binary parameters of BX Dra were initialized in a manner similar to that for the contact systems BX Peg (Lee et al. 2004) and
AA UMa (Lee et al. 2011). The effective temperature of the more massive primary star was fixed at $T_{1}$ = 6,980 K, according to
the spectral type F0 classified by Pych et al. (2004). Linear bolometric ($X$, $Y$) and monochromatic ($x$, $y$) limb-darkening coefficients
were interpolated from the values of van Hamme (1993) in concert with the model atmosphere option. The initial value for
the mass ratio ($q = m\rm_2/m\rm_1$) was taken from Pych et al. (2004). Throughout the analyses, synchronous rotation was assumed and
a third light source ($\ell_3$) was considered. In addition, as the atmosphere of both components should lie close to the boundary
between the radiative (hereafter RE) and convective (hereafter CE) envelopes, the gravity-darkening exponents ($g$) and
the bolometric albedos ($A$) were investigated at standard values of ($A$, $g$) = (1.0, 1.0) and (0.5, 0.32) for the two cases, respectively.
Because the CE model gives a better fit than does the RE model, the common envelope was treated as a convective atmosphere in
all subsequent syntheses. In Tables 3 and 4, parentheses represent adjusted parameters and the subscripts 1 and 2 refer to the primary
and secondary stars being eclipsed at Min I and Min II, respectively.

First of all, we analyzed simultaneously all the light curves of BX Dra, permitting no spots. The unspotted solution is listed in
the second column of Table 3 and the $V$ residuals from the analysis are plotted in the left panels of Figure 3. Similar patterns exist
for the other bandpasses. As shown in these panels, the model light curves do not fit the observed ones at all well. In contact binaries,
the discrepancies may be caused by local photospheric inhomogeneities such as a cool spot on a magnetically active component and/or
a hot spot due to impact from a mass transfer between the components (see, e.g., Lee et al. 2009b, 2010). Thus, spot models were added
to fit the light variations. Because the seasonal variations of the light residuals are not large, we reanalyzed the datasets considering
each of a hot and cool spot on each component. The results are given in columns (3)--(6) of Table 3 together with the spot parameters.
The residuals from the hot-spot model (Hot 1) on the primary component are plotted in the right panels of Figure 3. From these results,
we can see that the Hot 1 model gives slightly smaller values for the sum of the residuals squared ($\Sigma W(O-C)^2$) than do
the other models and that a large hot spot on the primary star has been sufficient for the light-curve representations of BX Dra.
Nonetheless, it is difficult to distinguish between the spot models from only the light-curve analysis because the differences
among them are small.

Finally, to understand the long-term spot behaviors in detail, we solved five historical data sets separately using the hot-spot model
on the primary star. For this procedure, we adjusted only the orbital ephemeris, spot, and luminosity parameters among
the Hot 1 model parameters. The final results are given in Table 4 and the normalized $V$ observations with the model light curves are
plotted in Figure 4. The light ratios have not changed as a result of the spot modeling and the seasonal light curves could be represented
by a slightly variable hot spot. Because the colatitudes and longitudes of the spot have been almost constant with time and
the components of BX Dra should have shallow convective shells and at most weak magnetic activity as surmised from their temperatures,
it is possible to regard the main cause of the activity as a sporadically-variable mass transfer from the secondary to the primary component.
However, the streaming gas might usually lead to a hot spot around the primary equator, rather than the modeled colatitude very different
from 90$^\circ$. In reality, the intrinsic light variations of BX Dra could be caused by the simultaneous existence of a hot spot due to
mass transfer and a magnetic spot on a component, but separate trials for such spot configurations did not give a better fit than
the single hot-spot model. In all procedures, we looked for a possible third light source ($\ell_{3}$) as suggested by SGG but found
that the parameter remains indistinguishable from zero within its error.

In order to measure the physical properties of the binary system, we analyzed the radial-velocity curves of Pych et al. (2004)
with our light-curve parameters for the Hot 1 model. From the photometric and spectroscopic results, the absolute dimensions of BX Dra were
determined and listed in Table 5, where the radii ($R$) are the mean-volume radii evaluated using tabulations of Mochnacki (1984).
The luminosity ($L$) and bolometric magnitudes ($M_{\rm bol}$) were computed by adopting $T_{\rm eff}$$_\odot$ = 5,780 K and
$M_{\rm bol}$$_\odot$ = +4.73 for solar values. To estimate the uncertainty in the luminosity, it was assumed that the temperature of
each component have an error of 200 K in accordance with the unreliability in the spectral classification.
For the absolute visual magnitudes ($M_{\rm V}$), we used the bolometric corrections (BCs) from the scaling between $\log T$ and BC
given by Torres (2010). The absolute parameters presented in this paper are consistent with those of SGG within the uncertainties.

\section{Orbital Period Study}

From the SOAO observations, 12 weighted times of minimum light and their errors were determined using the method of Kwee \& van Woerden (1956).
In addition to these, eight and two eclipse timings were newly determined from the individual measurements of KIM and ZOLA, respectively.
Including our measurements, a total of 87 timings (36 photographic, 9 photoelectric, and 42 CCD) have been collected from the database of
Kreiner et al. (2001) and from more recent literature. All available photoelectric and CCD timings are listed in Table 6. Because many timings
of the system have been published with no errors, the following standard deviations were assigned to timing residuals based on
the observational method: $\pm$0.0316 d for photographic plate, $\pm$0.0019 d for photoelectric and $\pm$0.0012 d for CCD minima.
Relative weights for the period analysis of BX Dra were then scaled from the inverse squares of these values (Lee et al. 2007).

As mentioned in the Introduction, SGG reported that the period change of BX Dra could be represented by a parabolic variation.
After testing several other forms including this possibility, we found that the eclipse timings display a sinusoidal variation superposed
on an upward parabola, rather than varying in a monotonic pattern. Using the PERIOD04 program (Lenz \& Breger 2005), we looked to see
if the residuals from the quadratic fit represent real and periodic variations. As shown in the small box of the top panel of Figure 5,
a frequency of $f$ = 0.0000636 cycle d$^{-1}$ was detected corresponding to about 25 yr. The periodic variation suggests
a light-travel time (LTT) effect driven by the existence of a third component orbiting the eclipsing binary. Thus, the complete timing data
were fitted to the following quadratic {\it plus} LTT ephemeris:
\begin{eqnarray}
C = T_0 + PE + AE^2 + \tau_{3},
\end{eqnarray}
where $\tau_{3}$ is the LTT due to a circumbinary object in the system (Irwin 1952, 1959) and includes five parameters
($a_{12}\sin i_3$, $e$, $\omega$, $n$ and $T$). The Levenberg-Marquardt algorithm (Press et al. 1992) was applied to solve for
the eight parameters of the ephemeris and the results are summarized in Table 7, together with related quantities.
Our absolute dimensions given in Table 5 have been used for these and subsequent calculations.

The $O$--$C$ residuals calculated from the linear terms in equation (1) are plotted in the top panel of Figure 5,
where the continuous curve represents the quadratic term of this ephemeris. The middle panel displays the LTT orbit, and
the bottom panel shows the residuals from the complete ephemeris. These appear as $O$--$C_{\rm full}$ in the fourth column of Table 6.
As displayed in the figure, the quadratic {\it plus} LTT ephemeris currently provides a good representation of all modern times of
minimum light. The LTT orbit has a period of $P_3$ = 30.2 yr, a semi-amplitude of $K$ = 0.0062 d, a projected orbital semi-major axis
of $a_{12} \sin i_3$ = 1.08 AU, and an eccentricity of $e$ = 0.35. The mass function of the circumbinary object becomes
$f_3 (M_{3})$ = 0.00138 M$_\odot$ and its minimum mass is 0.23 M$_\odot$. Because only about 62 \% of the 30-year period has been
covered by the photoelectric and CCD data, future accurate timings are required to identify and understand the LTT effect.

As in the case of the contact binary AR Boo (Lee et al. 2009b) with a convective envelope, the quasi-sinusoidal period variation with
small amplitudes may be produced by asymmetrical eclipse minima due to spot activity (Kalimeris et al. 2002) and/or the method of measuring
the timings (Maceroni \& van't Veer 1994). The light-curve synthesis method gives more and better information with respect to the other methods,
which do not consider spot activity and are based only on observations during minima (Lee et al. 2009b). Because five datasets of BX Dra
were modeled for Hot 1 spot parameters, we calculated a minimum epoch for each eclipse in these datasets with the W-D code by means of
adjusting only the ephemeris epoch ($T_0$). The results are listed in Table 8 together with the previously tabulated timings for comparison
and are illustrated with the `x' symbols in Figure 5. The differences between the published minima and those obtained by the synthesis method
are significantly smaller than the observed amplitude (0.012 d) of the LTT variation. As shown in Figure 5, the light-curve timings agree
with our analysis of the $O$--$C$ diagram and the periodic variation cannot result from the starspot activity. Further, as shown in
the fourth column of Table 8, there are systematic runs of differences between them, which are negative for Min I and positive for Min II.
These differences are caused by the hot spot on the primary star presented to the observer.

The quadratic term (A) of equation (1) signifies a continuous period increase with a rate of $+$(5.65$\pm$0.07)$\times$10$^{-7}$ d yr$^{-1}$,
corresponding to a fractional period change of $+$(1.55$\pm$0.02)$\times$10$^{-9}$. This is very close to the value of $+$1.29$\times$10$^{-9}$
calculated from the W-D code, independently of the eclipse timings. The most common explanation of the period increase is a mass transfer
from the less massive secondary star to the primary component. Under the assumption of conservative mass transfer, the transfer rate is
calculated to be 2.74$\times$10$^{-7}$ M$_\odot$ yr$^{-1}$. This value is larger than those recently derived for other overcontact systems
[e.g., 2.0$\times$10$^{-8}$ for BV Dra (Yang et al. 2009), 1.5$\times$10$^{-7}$ for AR Boo (Lee et al. 2009b), 7.4$\times$10$^{-8}$ for
V1191 Cyg (Zhu et al. 2011), and 6.6$\times$10$^{-8}$ for AA UMa (Lee et al. 2011), all in M$_\odot$ yr$^{-1}$ units].

\section{Summary And Discussion}

In this article, we have presented the long-term photometric behaviors of BX Dra from the detailed analyses of the light curves and
the $O$--$C$ diagram based on the historical and new observations. The results from these analyses can be summarized as follows:

\begin{enumerate}
\item Historical light curves of BX Dra, as well as our own, display total eclipses at secondary minima and inverse O'Connell effects
with Max I fainter than Max II. The asymmetric light curves can satisfactorily be explained by the spot model.
The slightly variable hot spot on the primary star permits good light-curve representations for the system.

\item The orbital period of BX Dra has varied in a combination with an upward parabola and a sinusoidal variation with a period of
30.2 yrs and a semi-amplitude of 0.0062 d, rather than in a monotonic fashion. The increasing rate of the secular period is calculated
to be $+$5.65$\times$10$^{-7}$ d yr$^{-1}$, which may be caused by the mass transfer from the secondary to the primary component in
the system with a rate of about 2.74$\times$10$^{-7}$ M$_\odot$ yr$^{-1}$.

\item The sinusoidal variation can be interpreted as the LTT effect due to the existence of a circumbinary object. If the third companion
is on the main sequence and its orbit is coplanar with the eclipsing pair ($i_3$ = 81$^\circ$.8), the mass of the circumbinary object
is $M_3$ = 0.23 M$_\odot$ and its radius and temperature are calculated to be $R_3$ = 0.24 R$_\odot$ and $T_3$ = 3022 K, respectively,
following the empirical relations from Southworth (2009). These correspond to a spectral type of about M6 V and a bolometric luminosity
of $L_3$ = 0.004 L$_\odot$ and contribute about 0.03\% to the total light of the triple system. So, it would be difficult to detect
such a faint companion from analyses of spotted light curves and spectroscopic observations. The absence of the evidence does not rule out
the presence of the hypothetic third component.

\item The results presented in this paper indicate that BX Dra is an A-subtype overcontact binary with an unseen circumbinary object;
the more massive primary star has a spectral type of F0 and the secondary component a spectral type of F1-2. We think that the hot spot
on the primary star could be produced as a result of the impact of the gas stream from the less massive secondary star.
The higher mass transfer rate than the other overcontact systems and no variation of spot parameters for about four years support
this hot spot model. Moreover, because both components should not have deep convective envelopes as surmised from their temperatures,
it is not reasonable to imagine that magnetic spots on the components of the system is the dominant cause of the light variation.
\end{enumerate}

Although all historical timings of BX Dra are in good agreement with the calculated LTT effect as seen in Figure 5, because of
the absence of any independent third body detection, we consider the possibility that a magnetic activity cycle may be the main cause
of the sinusoidal period modulation (Applegate 1992, Lanza et al. 1998). With the period ($P_3$) and amplitudes ($K$) listed in
Table 7, the model parameters were calculated for each component from the Applegate formulae. The parameters are listed in Table 9,
where the rms luminosity changes ($\Delta m_{\rm rms}$) converted to magnitude scale were obtained with equation (4) in the paper
of Kim et al. (1997). The variations of the gravitational quadrupole moment ($\Delta Q$) are two orders of magnitude smaller than
typical values of $10^{51}\sim10^{52}$ for close binaries (Lanza \& Rodono 1999). In reality, the magnetic mechanism is adequate
for systems with a spectral type later than about F5 (Hall 1989), unlike the case of BX Dra. Moreover, it is difficult for this model
to produce perfectly smooth and tilted periodic components in the $O$--$C$ variation. At present, because there is no other alternative
but the LTT effect, the sinusoidal variation most likely arises from an unseen third companion gravitationally bound to the eclipsing binary.
The circumbinary object in BX Dra may have played an important role in the formation of the eclipsing pair, which would finally evolve
into single star by angular momentum loss through magnetic braking. Future high-precision timing measurements will be crucial unveiling
the orbital period change of this system.

\bigskip

We would like to thank the staff of the Sobaeksan Optical Astronomy Observatory for assistance during our observations. We also thanks
F. Agerer, F. S\'anchez-Bajo, and S. Zola for sending us their published data on BX Dra. This research has made use of the Simbad database
maintained at CDS, Strasbourg, France. This work was supported by the KASI (Korea Astronomy and Space Science Institute) grant.

\clearpage
\begin{figure}
  \begin{center}
    \FigureFile(150mm,150mm){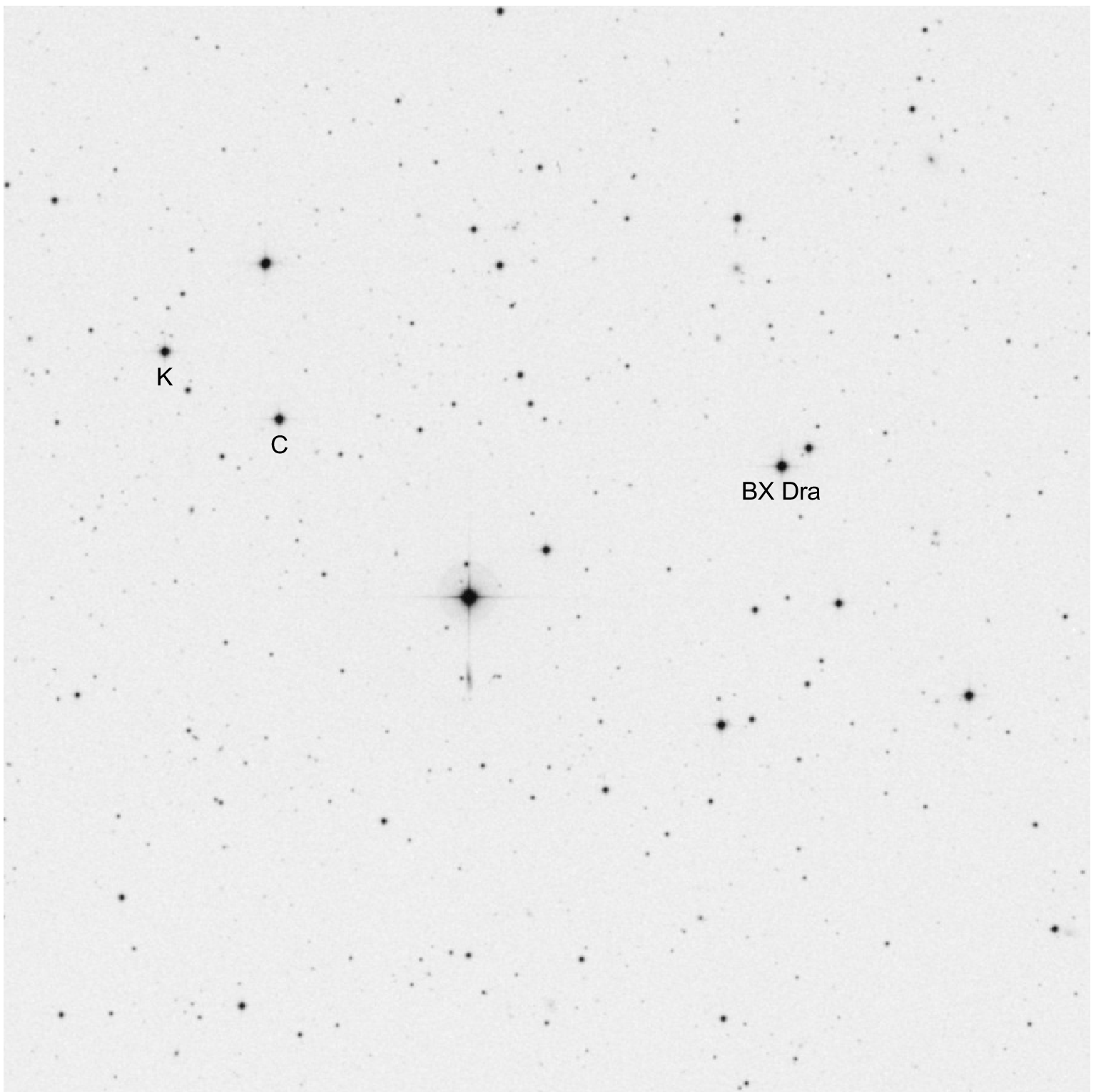}
  \end{center}
  \caption{A sample of observed CCD images (20$\arcmin$.5$\times$20\arcmin.5) of BX Dra from the 2009 season. Monitoring numerous frames
 led us to choose star C as a comparison and K as a check. North is up and east is to the left.}
  \label{Fig1}
\end{figure}

\begin{figure}
  \begin{center}
    \FigureFile(150mm,150mm){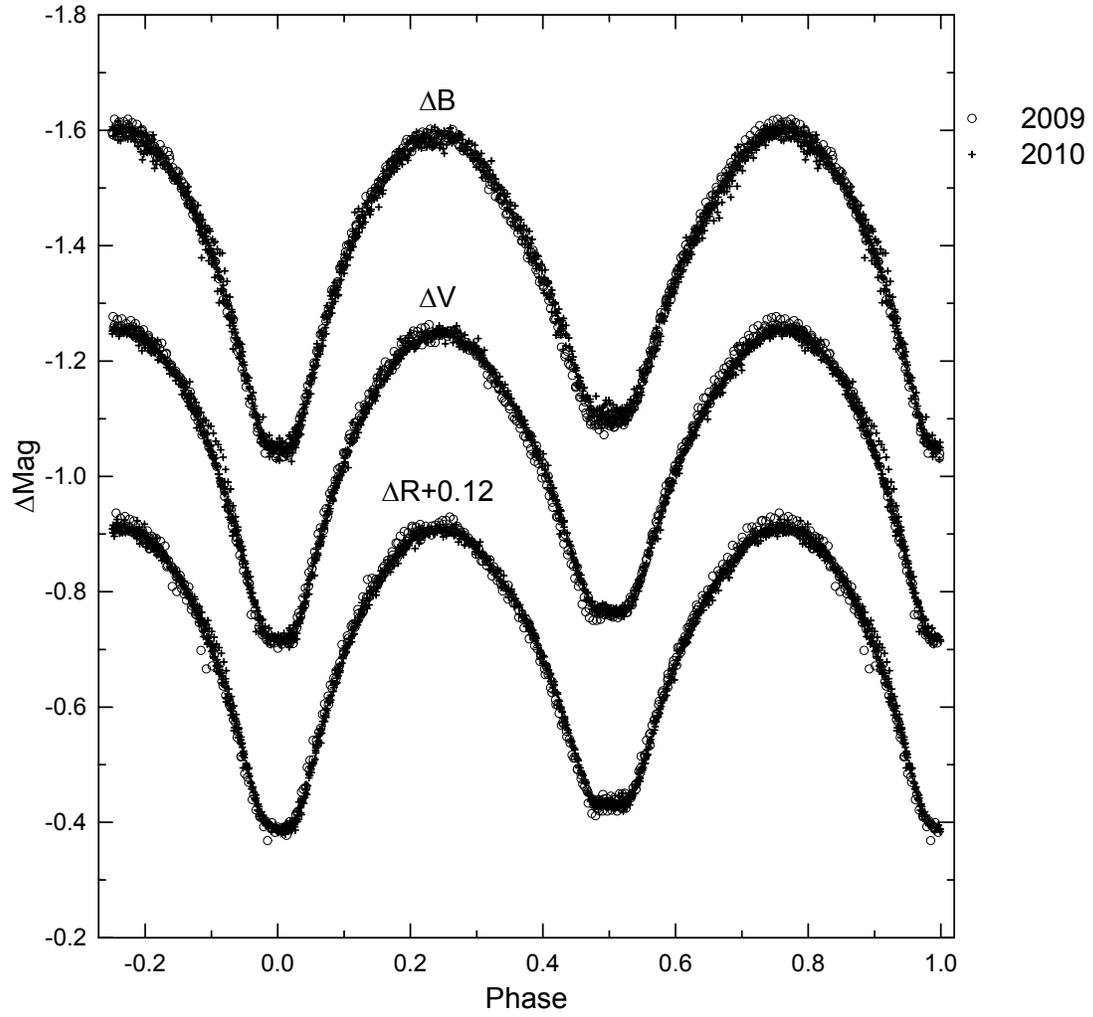}
  \end{center}
  \caption{$BVR$ light curves of BX Dra observed in 2009 and 2010. Because of the high density of the points, many of
 the 2009 measures cannot be seen individually.}
  \label{Fig2}
\end{figure}

\begin{figure}
  \begin{center}
    \FigureFile(150mm,150mm){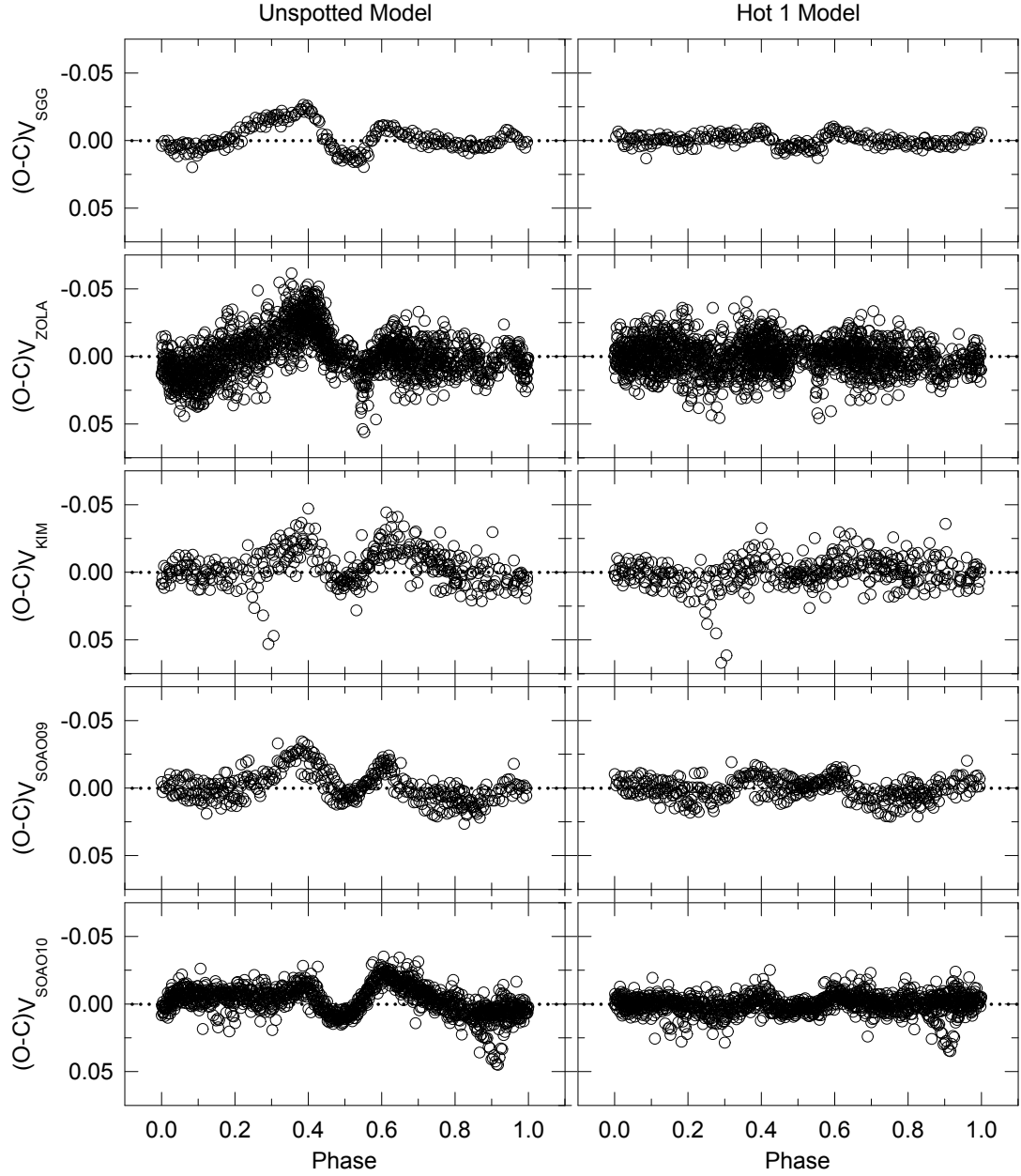}
  \end{center}
  \caption{Left and right panels show the light residuals from the solutions without a spot and with a hot spot (Hot 1) listed in Table 3,
 respectively. The data refer to the $V$ bandpass.}
  \label{Fig3}
\end{figure}

\begin{figure}
  \begin{center}
    \FigureFile(150mm,150mm){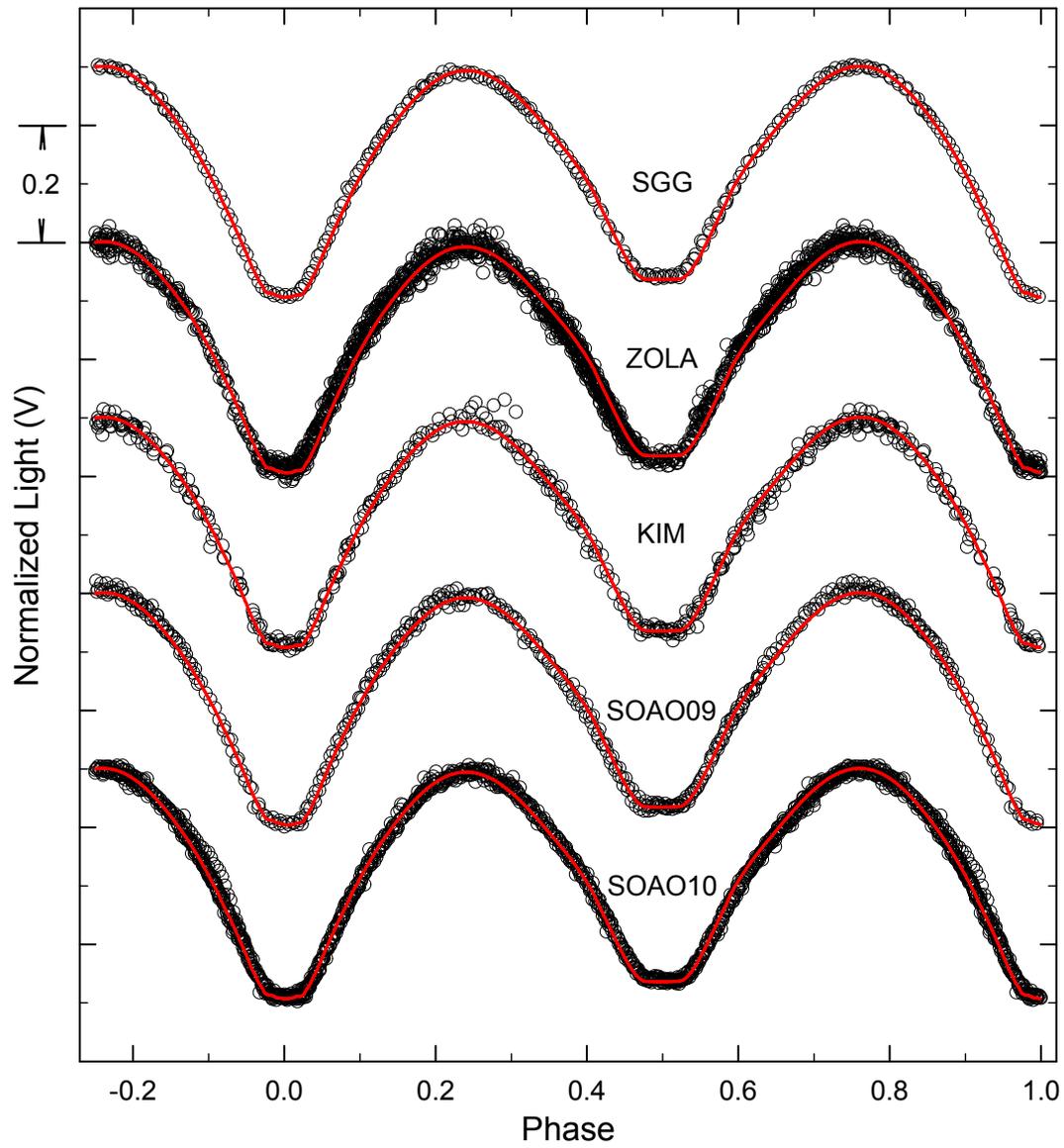}
  \end{center}
  \caption{Normalized $V$ observations with the fitted model light curves. The continuous curves represent the solutions obtained
 with the hot-spot model parameters listed in Table 4.}
  \label{Fig4}
\end{figure}

\begin{figure}
  \begin{center}
    \FigureFile(150mm,150mm){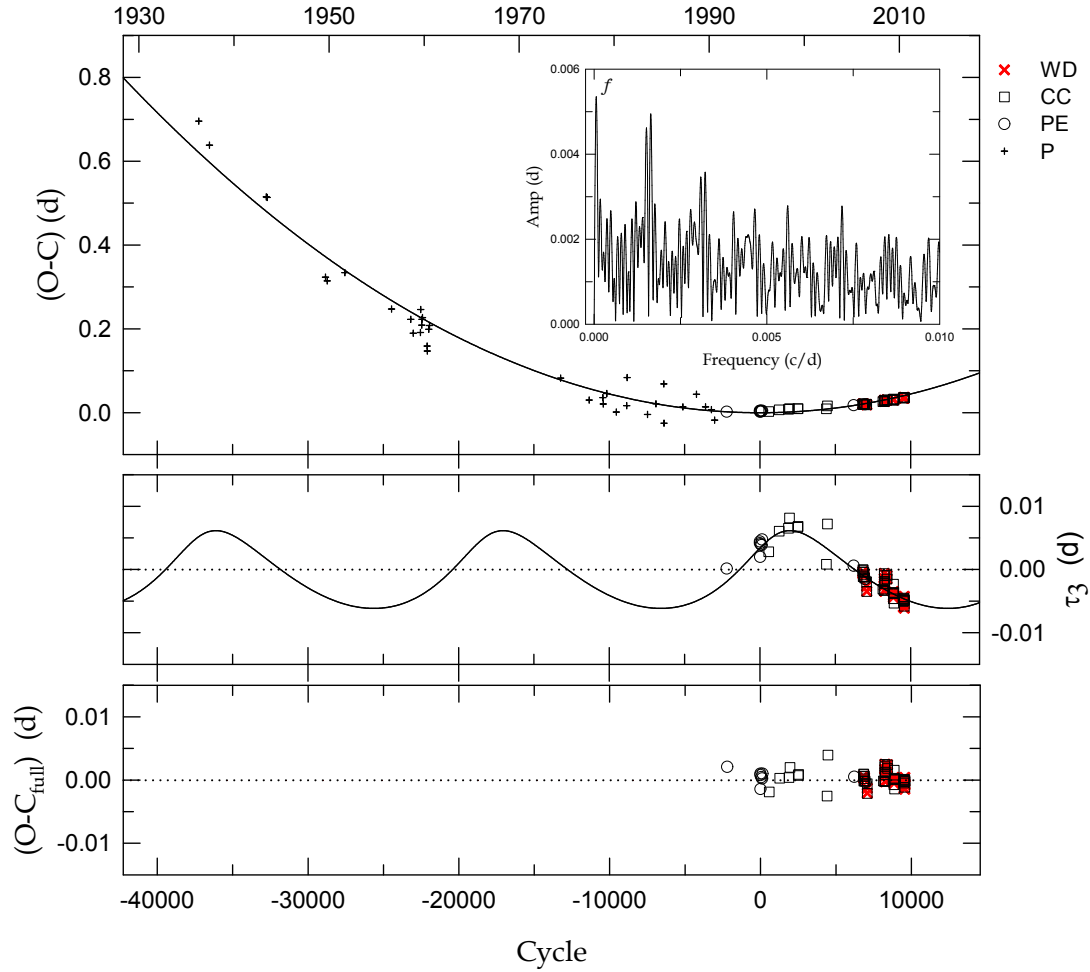}
  \end{center}
  \caption{$O$--$C$ diagram of BX Dra. In the top panel, constructed with the linear terms of the quadratic {\it plus} LTT ephemeris,
 the continuous curve represents the quadratic term of the equation. The inset box is drawn with the result from  the PERIOD04 formalism
 for the residuals, where a dominant frequency of $f$ = 0.0000636 cycle d$^{-1}$ is found with a semi-amplitude of 0.0054 mag.
 The middle panel displays the LTT orbit and the bottom panel shows the CC and PE residuals from the complete ephemeris. The `x' symbols
 refer to new minimum timings obtained from the W-D code.}
  \label{Fig5}
\end{figure}

\clearpage
\begin{table}
\caption{CCD photometric observations of BX Dra in 2009 and 2010.$\rm ^*$}
\label{tab1}
\begin{center}
\begin{tabular}{crcrcr}
\hline\hline
HJD             &  $\Delta B$&  HJD              &  $\Delta V$&  HJD              &  $\Delta R$    \\ \hline
\multicolumn{6}{c}{2009}                                                                           \\
2,454,927.16950 &  $-$1.102  &  2,454,927.16762  &  $-$0.773  &  2,454,927.16621  &  $-$0.571      \\
2,454,927.17432 &  $-$1.108  &  2,454,927.17257  &  $-$0.758  &  2,454,927.17122  &  $-$0.558      \\
2,454,927.17906 &  $-$1.109  &  2,454,927.17731  &  $-$0.765  &  2,454,927.17597  &  $-$0.565      \\
2,454,927.18381 &  $-$1.102  &  2,454,927.18206  &  $-$0.759  &  2,454,927.18071  &  $-$0.572      \\
2,454,927.18856 &  $-$1.110  &  2,454,927.18680  &  $-$0.764  &  2,454,927.18545  &  $-$0.573      \\
          \dots &     \dots  &            \dots  &     \dots  &            \dots  &     \dots      \\
\multicolumn{6}{c}{2010}                                                                           \\
2,455,303.26231 &  $-$1.065  &  2,455,303.26331  &  $-$0.724  &  2,455,303.26406  &  $-$0.521      \\
2,455,303.26501 &  $-$1.077  &  2,455,303.26602  &  $-$0.742  &  2,455,303.26677  &  $-$0.529      \\
2,455,303.26770 &  $-$1.084  &  2,455,303.26872  &  $-$0.749  &  2,455,303.26946  &  $-$0.547      \\
2,455,303.27049 &  $-$1.105  &  2,455,303.27163  &  $-$0.774  &  2,455,303.27245  &  $-$0.558      \\
2,455,303.27351 &  $-$1.125  &  2,455,303.27465  &  $-$0.792  &  2,455,303.27548  &  $-$0.588      \\
          \dots &     \dots  &            \dots  &     \dots  &            \dots  &     \dots      \\ \hline
\multicolumn{6}{l}{$\rm ^*$ A sample is shown here: the full version is available in its entirety} \\
\multicolumn{6}{l}{in a machine-readable form in the PASJ online edition.} \\
\end{tabular}
\end{center}
\end{table}

\clearpage
\begin{table}
\caption{Light-curve sets for BX Dra.}
\label{tab2}
\begin{center}
\begin{tabular}{lccc}
\hline\hline
Reference   & Season      & Data Type   & $\sigma$$\rm ^a$      \\ \hline
SGG         & 2006        & $B$         & 0.0078                \\
            &             & $V$         & 0.0062                \\
            &             & $I$         & 0.0058                \\
ZOLA        & 2006        & $B$         & 0.0115                \\
            &             & $V$         & 0.0121                \\
            &             & $R$         & 0.0120                \\
            &             & $I$         & 0.0113                \\
KIM         & 2008        & $B$         & 0.0209                \\
            &             & $V$         & 0.0119                \\
SOAO        & 2009        & $B$         & 0.0074                \\
            &             & $V$         & 0.0087                \\
            &             & $R$         & 0.0093                \\
            & 2010        & $B$         & 0.0098                \\
            &             & $V$         & 0.0070                \\
            &             & $R$         & 0.0062                \\ \hline
\multicolumn{4}{l}{$^a$ In units of total light at phase 0.75.} \\
\end{tabular}
\end{center}
\end{table}

\clearpage
\begin{table}
\caption{Binary parameters obtained by fitting simultaneously all light curves.}
\label{tab3}
\begin{center}
\begin{tabular}{lccccc}
\hline\hline
Parameter                             & Without Spot     & \multicolumn{4}{c}{Spot Model$\rm ^a$}                                    \\  [1.0mm] \cline{3-6} \\ [-2.0ex]
                                      &                  & Cool 1           & Cool 2           & Hot 1            & Hot 2            \\ \hline
$T_0$ (HJD)$^b$                       & 810.58987(12)    & 810.59009(13)    & 810.58995(13)    & 810.58976(12)    & 810.58970(12)    \\
$P$ (d)                               & 0.57902455(4)    & 0.57902476(4)    & 0.57902477(4)    & 0.57902479(4)    & 0.57902472(4)    \\
$dP$/$dt$ ($\times 10^{-9}$)          & 1.386(12)        & 1.303(11)        & 1.302(11)        & 1.288(11)        & 1.317(11)        \\
$T_1$ (K)                             & 6980             & 6980             & 6980             & 6980             & 6980             \\
$T_2$ (K)                             & 6979(2)          & 6707(2)          & 6995(2)          & 6758(4)          & 6805(3)          \\
$i$ (deg)                             & 80.63(6)         & 80.66(5)         & 82.42(8)         & 81.80(6)         & 82.18(8)         \\
$q$ ( = $m\rm_2/m\rm_1$)              & 0.2884(5)        & 0.2874(5)        & 0.2871(4)        & 0.2884(5)        & 0.2882(5)        \\
$\Omega_1$ = $\Omega_2$               & 2.3475(16)       & 2.3462(15)       & 2.3404(14)       & 2.3308(14)       & 2.3293(16)       \\
$L_1$/($L_{1}$+$L_{2}$){$_{B}$}       & 0.7432(5)        & 0.7828(6)        & 0.7407(4)        & 0.7729(4)        & 0.7660(5)        \\
$L_1$/($L_{1}$+$L_{2}$){$_{V}$}       & 0.7435(5)        & 0.7744(5)        & 0.7416(4)        & 0.7660(4)        & 0.7606(4)        \\
$L_1$/($L_{1}$+$L_{2}$){$_{R}$}       & 0.7437(4)        & 0.7685(5)        & 0.7421(4)        & 0.7612(4)        & 0.7568(4)        \\
$L_1$/($L_{1}$+$L_{2}$){$_{I}$}       & 0.7437(4)        & 0.7639(4)        & 0.7424(4)        & 0.7575(3)        & 0.7539(4)        \\
$r_1$ (pole)                          & 0.4791(4)        & 0.4792(4)        & 0.4804(3)        & 0.4828(3)        & 0.4832(4)        \\
$r_1$ (side)                          & 0.5218(5)        & 0.5219(5)        & 0.5237(5)        & 0.5272(5)        & 0.5277(5)        \\
$r_1$ (back)                          & 0.5542(7)        & 0.5543(7)        & 0.5565(6)        & 0.5614(7)        & 0.5620(7)        \\
$r_2$ (pole)                          & 0.2803(6)        & 0.2799(6)        & 0.2811(5)        & 0.2846(6)        & 0.2849(6)        \\
$r_2$ (side)                          & 0.2953(8)        & 0.2949(7)        & 0.2964(7)        & 0.3007(8)        & 0.3010(8)        \\
$r_2$ (back)                          & 0.3513(18)       & 0.3506(17)       & 0.3542(16)       & 0.3638(20)       & 0.3648(21)       \\
\multicolumn{6}{l}{Spot parameters:}                                                                                                 \\
Colatitude$_1$ (deg)                  & \dots            & 22.3             & \dots            & 8.5              & \dots            \\
Longitude$_1$ (deg)                   & \dots            & 188.8            & \dots            & 11.5             & \dots            \\
Radius$_1$ (deg)                      & \dots            & 35.7             & \dots            & 36.5             & \dots            \\
$T$$\rm _{spot1}$/$T$$\rm _{local1}$  & \dots            & 0.758            & \dots            & 1.361            & \dots            \\
Colatitude$_2$ (deg)                  & \dots            & \dots            & 24.1             & \dots            & 33.3             \\
Longitude$_2$ (deg)                   & \dots            & \dots            & 9.6              & \dots            & 186.0            \\
Radius$_2$ (deg)                      & \dots            & \dots            & 50.4             & \dots            & 33.2             \\
$T$$\rm _{spot2}$/$T$$\rm _{local2}$  & \dots            & \dots            & 0.723            & \dots            & 1.259            \\
$\Sigma W(O-C)^2$                     & 0.0151           & 0.0129           & 0.0126           & 0.0121           & 0.0126           \\ \hline
\multicolumn{6}{l}{$^a$ Cool 1: a cool spot on the primary; Cool 2: a cool spot on the secondary; Hot 1: a hot spot} \\
\multicolumn{6}{l}{on the primary; Hot 2: a hot spot on the secondary.} \\
\multicolumn{6}{l}{$^b$ HJD 2,449,000 is suppressed.} \\
\end{tabular}
\end{center}
\end{table}

\clearpage
\begin{table}
\caption{Year-to-year variations of the spot and luminosity parameters.}
\label{tab4}
\begin{center}
\begin{tabular}{lccccc}
\hline\hline
Parameter                            & SGG                  & ZOLA                 & KIM                  & SOAO09               & SOAO10               \\ \hline
$T_0$ (HJD)$^a$                      & 3767.66365(7)        & 3905.47131(5)        & 4581.20182(19)       & 4932.09274(8)        &  5314.25365(7)       \\
$P$ (d)                              & 0.5790304(16)        & 0.5790455(73)        & 0.5790273(20)        & 0.5790446(45)        & 0.5790425(13)        \\
Colatitude$_1$ (deg)                 & 8.5                  & 8.5                  & 8.8                  & 8.5                  & 8.3                  \\
Longitude$_1$ (deg)                  & 10.9                 & 11.5                 & 11.5                 & 11.5                 & 11.5                 \\
Radius$_1$ (deg)                     & 36.1                 & 39.1                 & 38.5                 & 39.6                 & 36.5                 \\
$T$$\rm _{spot1}$/$T$$\rm _{local1}$ & 1.365                & 1.369                & 1.386                & 1.340                & 1.370                \\
$L_1$/($L_{1}$+$L_{2}$){$_{B}$}      & 0.7729(4)            & 0.7729(3)            & 0.7729(10)           & 0.7729(4)            & 0.7729(2)            \\
$L_1$/($L_{1}$+$L_{2}$){$_{V}$}      & 0.7660(3)            & 0.7660(3)            & 0.7660(5)            & 0.7660(3)            & 0.7660(2)            \\
$L_1$/($L_{1}$+$L_{2}$){$_{R}$}      & \dots                & 0.7612(3)            & \dots                & 0.7612(3)            & 0.7612(2)            \\
$L_1$/($L_{1}$+$L_{2}$){$_{I}$}      & 0.7575(2)            & 0.7575(3)            & \dots                & \dots                & \dots                \\
$\Sigma W(O-C)^2$                    & 0.0092               & 0.0119               & 0.0113               & 0.0121               & 0.0116               \\ \hline
\multicolumn{6}{l}{$^a$ HJD 2,450,000 is suppressed.} \\
\end{tabular}
\end{center}
\end{table}

\clearpage
\begin{table}
\caption{Absolute parameters for BX Dra.}
\label{tab5}
\begin{center}
\begin{tabular}{lccc}
\hline\hline
Parameter               & Primary           & Secondary    \\ \hline
$a$ ($R_\odot$)         & \multicolumn{2}{c}{4.058(87)}    \\
$V_0$ (km s$^{-1}$)     & \multicolumn{2}{c}{$-$25.6(2.2)} \\
$M$ ($M_\odot$)         & 2.08(10)          & 0.60(4)      \\
$R$ ($R_\odot$)         & 2.13(5)           & 1.28(3)      \\
log g (cgs)             & 4.10(3)           & 4.00(3)      \\
$\rho$ (g cm$^{-3}$)    & 0.30(2)           & 0.41(4)      \\
$T$ (K)                 & 6980(200)         & 6758(200)    \\
$L$ ($L_\odot$)         & 9.66(1.18)        & 3.05(38)     \\
$M_{\rm bol}$ (mag)     & 2.27(13)          & 3.52(14)     \\
BC (mag)                & 0.03              & 0.02         \\
$M_{\rm v}$ (mag)       & 2.24(13)          & 3.50(14)     \\ \hline
\end{tabular}
\end{center}
\end{table}

\clearpage
\begin{table}
\caption{Photoelectric and CCD timings of minimum light for BX Dra.}
\label{tab6}
\begin{center}
\begin{scriptsize}
\begin{tabular}{lcrrcl}
\hline\hline
HJD          & Error         & Epoch     & $O$--$C_{\rm full}$ & Min & References          \\
(2,400,000+) &               &           &                     &     &                     \\ \hline
48,528.63    &               & $-$2214.0 &    0.0021 & I  & Agerer \& Dahm (1995)          \\
49,810.5926  &               &       0.0 &    0.0008 & I  & Agerer \& Dahm (1995)          \\
49,811.4614  &               &       1.5 &    0.0010 & II & Agerer \& Dahm (1995)          \\
49,812.3275  &               &       3.0 & $-$0.0014 & I  & Agerer \& Dahm (1995)          \\
49,840.4122  &               &      51.5 &    0.0005 & II & Agerer \& Dahm (1995)          \\
49,866.4682  &               &      96.5 &    0.0002 & II & Agerer \& Dahm (1995)          \\
49,888.4720  &               &     134.5 &    0.0010 & II & Agerer \& Dahm (1995)          \\
50,147.5838  & $\pm$0.0003   &     582.0 & $-$0.0019 & I  & Agerer \& H\"ubscher (1996)    \\
50,547.4042  & $\pm$0.0004   &    1272.5 &    0.0003 & II & Agerer \& H\"ubscher (1999)    \\
50,904.3743  & $\pm$0.0003   &    1889.0 &    0.0004 & I  & Agerer \& H\"ubscher (1999)    \\
50,945.4868  & $\pm$0.0005   &    1960.0 &    0.0020 & I  & Agerer \& H\"ubscher (1999)    \\
51,256.4227  & $\pm$0.0004   &    2497.0 &    0.0007 & I  & Agerer \& H\"ubscher (2000)    \\
51,270.6090  & $\pm$0.0011   &    2521.5 &    0.0009 & II & Agerer \& H\"ubscher (2000)    \\
52,362.360   & $\pm$0.003    &    4407.0 & $-$0.0025 & I  & Diethelm (2002)                \\
52,401.4508  & $\pm$0.0008   &    4474.5 &    0.0039 & II & Agerer \& H\"ubscher (2003)    \\
53,409.5346  & $\pm$0.0007   &    6215.5 &    0.0005 & II & H\"ubscher et al. (2005)       \\
53,758.6891  & $\pm$0.0002   &    6818.5 &    0.0006 & II & S\'anchez-Bajo et al. (2007)   \\
53,767.6644  & $\pm$0.0003   &    6834.0 &    0.0010 & I  & S\'anchez-Bajo et al. (2007)   \\
53,772.5851  & $\pm$0.0001   &    6842.5 & $-$0.0002 & II & S\'anchez-Bajo et al. (2007)   \\
53,774.6127  & $\pm$0.0001   &    6846.0 &    0.0009 & I  & S\'anchez-Bajo et al. (2007)   \\
53,794.5882  & $\pm$0.0001   &    6880.5 & $-$0.0001 & II & S\'anchez-Bajo et al. (2007)   \\
53,800.6687  & $\pm$0.0001   &    6891.0 &    0.0006 & I  & S\'anchez-Bajo et al. (2007)   \\
53,803.5635  & $\pm$0.0001   &    6896.0 &    0.0002 & I  & S\'anchez-Bajo et al. (2007)   \\
53,829.9089  & $\pm$0.0002   &    6941.5 & $-$0.0002 & II & Nelson (2007)                  \\
53,846.4111  & $\pm$0.0004   &    6970.0 & $-$0.0003 & I  & Diethelm (2006)                \\
53,905.4719  & $\pm$0.0002   &    7072.0 & $-$0.0006 & I  & Zola et al. (2010)             \\
53,907.4969  & $\pm$0.0002   &    7075.5 & $-$0.0021 & II & Zola et al. (2010)             \\
54,527.9296  & $\pm$0.0001   &    8147.0 & $-$0.0001 & I  & Nelson (2009)                  \\
54,575.1205  & $\pm$0.0002   &    8228.5 & $-$0.0002 & II & Kim et al. (2009)              \\
54,581.2027  & $\pm$0.0003   &    8239.0 &    0.0021 & I  & Kim et al. (2009)              \\
54,584.0982  & $\pm$0.0006   &    8244.0 &    0.0025 & I  & Kim et al. (2009)              \\
54,586.1232  & $\pm$0.0007   &    8247.5 &    0.0009 & II & Kim et al. (2009)              \\
54,588.1501  & $\pm$0.0005   &    8251.0 &    0.0012 & I  & Kim et al. (2009)              \\
54,597.4133  & $\pm$0.0001   &    8267.0 & $-$0.0001 & I  & H\"ubscher (2009)              \\
54,603.2050  & $\pm$0.0006   &    8277.0 &    0.0013 & I  & Kim et al. (2009)              \\
54,685.1390  & $\pm$0.0003   &    8418.5 &    0.0024 & II & Kim et al. (2009)              \\
54,686.0070  & $\pm$0.0005   &    8420.0 &    0.0018 & I  & Kim et al. (2009)              \\
54,931.2241  & $\pm$0.0003   &    8843.5 & $-$0.0007 & II & This paper (SOAO)              \\
54,932.0935  & $\pm$0.0003   &    8845.0 &    0.0001 & I  & This paper (SOAO)              \\
54,934.1202  & $\pm$0.0003   &    8848.5 &    0.0002 & II & This paper (SOAO)              \\
54,941.3595  & $\pm$0.0003   &    8861.0 &    0.0016 & I  & Br\'at et al. (2009)           \\
54,951.2017  & $\pm$0.0002   &    8878.0 &    0.0003 & I  & This paper (SOAO)              \\
54,955.8323  & $\pm$0.0012   &    8886.0 & $-$0.0014 & I  & Diethelm (2009)                \\
55,314.2540  & $\pm$0.0004   &    9505.0 & $-$0.0005 & I  & This paper (SOAO)              \\
55,327.2826  & $\pm$0.0002   &    9527.5 & $-$0.0000 & II & This paper (SOAO)              \\
55,332.2046  & $\pm$0.0001   &    9536.0 &    0.0001 & I  & This paper (SOAO)              \\
55,337.1261  & $\pm$0.0002   &    9544.5 & $-$0.0002 & II & This paper (SOAO)              \\
55,350.1545  & $\pm$0.0002   &    9567.0 &    0.0000 & I  & This paper (SOAO)              \\
55,352.1809  & $\pm$0.0002   &    9570.5 & $-$0.0002 & II & This paper (SOAO)              \\
55,353.0485  & $\pm$0.0002   &    9572.0 & $-$0.0012 & I  & This paper (SOAO)              \\
55,354.2074  & $\pm$0.0003   &    9574.0 & $-$0.0002 & I  & This paper (SOAO)              \\ \hline
\end{tabular}
\end{scriptsize}
\end{center}
\end{table}

\clearpage
\begin{table}
\caption{Parameters for the quadratic {\it plus} LTT ephemeris of BX Dra.}
\label{tab7}
\begin{center}
\begin{tabular}{ccc}
\hline\hline
Parameter               &  Values                         &  Unit                  \\ \hline
$T_0$                   &  2,449,810.58844(36)            &  HJD                   \\
$P$                     &  0.579024741(46)                &  d                     \\
$A$                     &  4.476(56)$\times 10^{-10}$     &  d                     \\
$a_{12}\sin i_{3}$      &  1.08(13)                       &  AU                    \\
$\omega$                &  61.8(3.8)                      &  deg                   \\
$e$                     &  0.350(94)                      &                        \\
$n  $                   &  0.0326(13)                     &  deg d$^{-1}$          \\
$T$                     &  2,450,417(140)                 &  HJD                   \\
$P_{3}$                 &  30.2(1.2)                      &  yr                    \\
$K$                     &  0.00615(74)                    &  d                     \\
$f(M_{3})$              &  0.00138(17)                    &  $M_\odot$             \\
$M_3 \sin i_{3}$        &  0.23                           &  $M_\odot$             \\
$dP$/$dt$               &  5.647(70)$\times 10^{-7}$      &  d yr$^{-1}$           \\
$dM$/$dt$               &  2.741$\times 10^{-7}$          &  $M_\odot$ yr$^{-1}$   \\ \hline
\end{tabular}
\end{center}
\end{table}

\clearpage
\begin{table}
\caption{Minimum timings determined by the W-D code from individual eclipses of BX Dra.}
\label{tab8}
\begin{center}
\begin{tabular}{cccrccl}
\hline\hline
Observed$\rm^{a,b}$ & W-D$\rm^{b}$ & Error$\rm^{c}$ & Difference$\rm^{d}$ & Filter & Min & References \\ \hline
3,758.6891 & 3,758.68907 & $\pm$0.00023 &    0.00000 & $BV$   & II & S\'anchez-Bajo et al. (2007)     \\
3,767.6644 & 3,767.66375 & $\pm$0.00015 & $+$0.00067 & $BV$   & I  & S\'anchez-Bajo et al. (2007)     \\
3,772.5851 & 3,772.58546 & $\pm$0.00022 & $-$0.00041 & $BV$   & II & S\'anchez-Bajo et al. (2007)     \\
3,774.6127 & 3,774.61214 & $\pm$0.00016 & $+$0.00056 & $BV$   & I  & S\'anchez-Bajo et al. (2007)     \\
3,794.5882 & 3,794.58853 & $\pm$0.00014 & $-$0.00032 & $I$    & II & S\'anchez-Bajo et al. (2007)     \\
3,800.6687 & 3,800.66844 & $\pm$0.00013 & $+$0.00029 & $I$    & I  & S\'anchez-Bajo et al. (2007)     \\
3,803.5635 & 3,803.56325 & $\pm$0.00007 & $+$0.00023 & $I$    & I  & S\'anchez-Bajo et al. (2007)     \\
3,905.4719 & 3,905.47111 & $\pm$0.00011 & $+$0.00075 & $BVRI$ & I  & Zola et al. (2010)               \\
3,907.4969 & 3,907.49692 & $\pm$0.00014 &    0.00000 & $BVRI$ & II & Zola et al. (2010)               \\
4,575.1205 & 4,575.12128 & $\pm$0.00047 & $-$0.00078 & $BV$   & II & Kim et al. (2009)                \\
4,581.2027 & 4,581.20221 & $\pm$0.00042 & $+$0.00049 & $BV$   & I  & Kim et al. (2009)                \\
4,584.0982 & 4,584.09820 & $\pm$0.00047 &    0.00000 & $BV$   & I  & Kim et al. (2009)                \\
4,586.1232 & 4,586.12320 & $\pm$0.00076 &    0.00000 & $BV$   & II & Kim et al. (2009)                \\
4,588.1501 & 4,588.14872 & $\pm$0.00053 & $+$0.00138 & $BV$   & I  & Kim et al. (2009)                \\
4,603.2050 & 4,603.20341 & $\pm$0.00055 & $+$0.00159 & $BV$   & I  & Kim et al. (2009)                \\
4,685.1390 & 4,685.13900 & $\pm$0.00047 &    0.00000 & $BV$   & II & Kim et al. (2009)                \\
4,686.0070 & 4,686.00700 & $\pm$0.00070 &    0.00000 & $BV$   & I  & Kim et al. (2009)                \\
4,931.2241 & 4,931.22464 & $\pm$0.00028 & $-$0.00051 & $BVR$  & II & This article                     \\
4,932.0935 & 4,932.09287 & $\pm$0.00018 & $+$0.00065 & $BVR$  & I  & This article                     \\
4,934.1202 & 4,934.12040 & $\pm$0.00014 & $-$0.00019 & $BVR$  & II & This article                     \\
4,951.2017 & 4,951.20127 & $\pm$0.00018 & $+$0.00047 & $BVR$  & I  & This article                     \\
5,314.2540 & 5,314.25309 & $\pm$0.00017 & $+$0.00089 & $BVR$  & I  & This article                     \\
5,327.2826 & 5,327.28296 & $\pm$0.00018 & $-$0.00032 & $BVR$  & II & This article                     \\
5,332.2046 & 5,332.20401 & $\pm$0.00009 & $+$0.00056 & $BVR$  & I  & This article                     \\
5,337.1261 & 5,337.12605 & $\pm$0.00018 &    0.00000 & $BVR$  & II & This article                     \\
5,350.1545 & 5,350.15382 & $\pm$0.00011 & $+$0.00063 & $BVR$  & I  & This article                     \\
5,352.1809 & 5,352.18145 & $\pm$0.00012 & $-$0.00058 & $BVR$  & II & This article                     \\
5,353.0485 & 5,353.04813 & $\pm$0.00013 & $+$0.00033 & $BVR$  & I  & This article                     \\
5,354.2074 & 5,354.20687 & $\pm$0.00020 & $+$0.00056 & $BVR$  & I  & This article                     \\ \hline
\multicolumn{6}{l}{$^a$ cf. Table 6.} \\
\multicolumn{6}{l}{$^b$ HJD 2,450,000 is suppressed.} \\
\multicolumn{6}{l}{$^c$ Uncertainties yielded by the W-D code.} \\
\multicolumn{6}{l}{$^d$ Differences between columns (1) and (2).} \\
\end{tabular}
\end{center}
\end{table}

\clearpage
\begin{table}
\caption{Applegate parameters for possible magnetic activity of BX Dra.}
\label{tab9}
\begin{center}
\begin{tabular}{cccc}
\hline\hline
Parameter                 &  Primary               & Secondary               & Unit                 \\ \hline
$\Delta P$                & 0.1753                 &  0.1753                 &  s                   \\
$\Delta P/P$              & $3.50\times10^{-6}$    &  $3.50\times10^{-6}$    &                      \\
$\Delta Q$                & ${1.28\times10^{50}}$  &  ${3.70\times10^{49}}$  &  g cm$^2$            \\
$\Delta J$                & ${2.60\times10^{47}}$  &  ${1.00\times10^{47}}$  &  g cm$^{2}$ s$^{-1}$ \\
$I_{\rm s}$               & ${6.05\times10^{54}}$  &  ${6.30\times10^{53}}$  &  g cm$^{2}$          \\
$\Delta \Omega$           & ${4.30\times10^{-8}}$  &  ${1.58\times10^{-7}}$  &  s$^{-1}$            \\
$\Delta \Omega / \Omega$  & ${3.42\times10^{-4}}$  &  ${1.26\times10^{-3}}$  &                      \\
$\Delta E$                & ${2.24\times10^{40}}$  &  ${3.15\times10^{40}}$  &  erg                 \\
$\Delta L_{\rm rms}$      & ${7.37\times10^{31}}$  &  ${1.04\times10^{32}}$  &  erg s$^{-1}$        \\
                          & 0.0192                 &  0.0271                 &  $L_\odot$           \\
                          & 0.0020                 &  0.0089                 &  $L_{1,2}$           \\
$\Delta m_{\rm rms}$      & $\pm$0.0016            &  $\pm$0.0023            &  mag                 \\
$B$                       & 4.1                    &  5.4                    &  kG                  \\ \hline
\end{tabular}
\end{center}
\end{table}

\end{document}